\documentstyle[osa,manuscript]{revtex}

\def\ee{e$^+$e$^-\;$}

\begin{document}

\title{Strangeness production in a constituent quark model}

\author{F.Becattini and G.Pettini$^{*}$}

\address{Department of Physics, University of Florence and I.N.F.N.,
Florence, Italy\\
E-mail: becattini@fi.infn.it, pettini@fi.infn.it}

\maketitle

\begin{abstract}
We develop a model to calculate strangeness production in both
elementary and heavy ion collisions, within the framework of a
statistical approach to hadronisation. Calculations are based on
the canonical partition function of the thermal Nambu-Jona-Lasinio
model with exact conservation of flavor and color. It turns out
that the growth of strange quarks production in heavy ion
collisions is due to the initial excess of non-strange matter over
antimatter, whereas a suppression occurs for elementary
collisions, owing to the constraint of exact quantum charges
conservation over small volumes.
\end{abstract}

\section{The Statistical Model of Hadronisation (SHM) and
the Wroblewski factor $\lambda_{S}$} \label{sec:shm}

A statistical calculation of hadron multiplicities in high energy
collisions has been developed in ref.~\cite{beca}. The SHM relies
on the assumption of equiprobability of multi-hadronic states
originating from hadronising clusters, entailing that particle
production in hadronisation of both elementary (EC) and heavy ion
(HIC) collisions can be treated as an equilibrium process. The
main advantage of the model is the low number of free parameters
necessary to reproduce the observed hadron multiplicities. These
are: the total volume $V$ of the set of clusters, temperature $T$,
and a phenomenological parameter $\gamma_S$ reducing the
production of strange particles with respect to a fully chemically
equilibrated hadron gas. The exact conservation of initial
electric, baryon and strange total quantum charges is enforced in
the model. However, when dealing with HIC, the conservation of the
initial baryon number can be imposed only on the average and the
baryon chemical potential $\mu_B$ has to be introduced. A
remarkable outcome of the SHM is that the temperature fitted in
any EC is almost constant ($T \sim 160$ MeV), and very close to
that obtained in HIC with small fitted $\mu_B$. Furthermore, for
different HIC, to an increase of the fitted value of $\mu_B$
corresponds a decrease of the value of $T$.

 An interesting insight in strangeness production is obtained
when studying the so-called Wroblewski factor, the ratio between
newly produced valence strange and light quark pairs
$\lambda_S=2\langle s {\bar s}\rangle/(\langle u{\bar
u}\rangle+\langle d{\bar d}\rangle)$, which is fairly constant in
several EC, with $\sqrt{s}$ spanning two orders of magnitude,
whereas it is significantly higher and more variable in HIC
~\cite{beca1} (see Fig.~\ref{fig:ls}).

The peculiar behaviour of $\lambda_S$, together with the apparent
universality of the scale of temperature fitted within EC, suggest
to look for a more fundamental description of strangeness
production, by employing effective models  (EM) having quarks as
fundamental degrees of freedom.

\section{Connection with effective models}\label{sec:em}

In using EM, the same physical scheme of the SHM is kept. However,
we now assume the formation of a set of hadronising clusters
endowed with given $U,D,S$ and color charges, in which every
allowed quantum state is equally likely. The assumption of
suitable maximum disorder fluctuations of cluster flavor charges
is retained ~\cite{beca1}, which allows to impose exact flavor
conservation over the total volume $V$. Further conjectures are
introduced:
\begin{enumerate}
\item{}Temperature $T$ and baryon chemical potential
$\mu_B$, fitted within the SHM, are interpreted as the critical
values for deconfinement and (approximate) chiral symmetry
restoration
\item{}The produced s quarks, or at least the ratio s/u, survive in the
hadronic phase
\item{}Each single cluster is a color singlet
\end{enumerate}


The first assumption is the strongest one, as it implies that
hadronisation itself is assumed to be a critical process. The
aforementioned  constancy of fitted temperature for many
collisions ~\cite{beca1} and the consistency with other estimates,
support this identification. Assumption (2) implies that full
chemical equilibrium is assumed at constituent quark level. Thus,
no parameter such as $\gamma_S$ is now required to account for
strangeness production. Conversely, assumption (3) introduces a
further parameter, a characteristic cluster size $V_c$ over which
color is exactly conserved, which, together with the constancy of
$T$ and a weak dependence of $\lambda_S$ itself on the total
volume $V$, contributes in stabilizing $\lambda_S$ for EC. On the
other hand, a large $V_c$ in HIC could be a signal for color
deconfinement over large volumes.

The simplest effective model to start with is the thermal
Nambu-Jona-Lasinio model ~\cite{njl}. It cannot account for
deconfinement but it does embody chiral symmetry breaking and its
restoration ($\chi$SR), which is expected to occur at the same
critical point ~\cite{karsch}. Although not renormalizable, the
NJL model shares the main features with other effective models,
referred in literature as {\em ladder}-QCD ~\cite{bcd}. Actually
the phase diagram for $\chi$SR exhibits a tricritical point in the
chiral limit, separating second order from first order phase
transitions. We are interested to include the current quark
masses, in which case a smooth cross-over transition is expected
for the light quarks in the second order region (in which we are
interested in the present work). This implies a quasi-critical
behavior of the light quark condensates and a smooth decrease of
the strange one for higher temperatures. The same happens to the
constituent quark masses which are the physical quantities,
together with the UV cutoff $\Lambda$ ~\cite{njl}, affecting the
number of quarks (or antiquarks). Actually, the thermodynamics of
the mean-field NJL model can be derived by a free Dirac
Hamiltonian having constituent quark masses replacing the current
ones. The dependence of quark multiplicities on $V,V_c$ emerges
when requiring exact conservation of quantum flavor and color
charges by restricting the partition function to the allowed set
of multi-particle states. This is accomplished by means of
standard methods ~\cite{exact} and implies a five-dimensional
integration over the group $SU(3)_c\times U(1)_u\times
U(1)_d\times U(1)_s$. Calculations are shown in detail in ref.
~\cite{noi}.
\section{Results}\label{sec:resu}
\subsection{Heavy Ions}\label{subs:hic}
For HIC , the fitted total volumes ~\cite{beca1} are large enough
to disregard the fluctuations of flavor quantum numbers. Then, if
$V_c$ is large enough, it can be explicitly checked that a
grand-canonical calculation of quark multiplicities is reliable.
This implies that $\lambda_S$ can be obtained by the simple
grand-canonical formula for a gas of free constituent quarks with
an UV cutoff limiting the momentum integration ~\cite{njl}, and it
is independent on $V,V_c$. Temperature $T$ and chemical potentials
$\mu_i$ are taken coincident with those fitted within the SHM. A
first comparison with the SHM predictions can be performed by
considering the values of the constituent quark masses $M_{i}$ for
the strange quark ($\mu_{s}=0$) and for a light quark $q$
$(\mu_{q}\sim \mu_{B}/3$) with mass ${\hat m}=(m_{u}+m_{d})/2$, as
free parameters to be determined imposing the SHM value of
$\lambda_S$ for a given process. A more complete analysis can be
obtained by using the thermal NJL model in a predictive way. The
coincidence of the cross-over point with the best fit of the SHM
is possible by tuning the parameter $T_0$, defined in
ref.~\cite{njl}, which controls the strength of the six-fermion
term and which physically represents the critical temperature for
$U(1)_{A}$ restoration. Then the cross-over line, obtained by
studying the chiral light quark susceptibility can be compared
with the fitted values of $T$ and $\mu_B$ in the SHM, as well as
$\lambda_S$ along the critical line. In this case the constituent
masses $M_{i}(T,\mu_{i})$ are predicted by minimizing the NJL
grand canonical potential. A good agreement is found with
$T_0=$170 MeV for various HIC.

However, as it turns out that $M_s$ is almost independent on the
choice of $T_0$, the model needs the specification of only one
parameter, either $T_0$ or $M_q$. This is shown in
Fig.~\ref{bande}, which shows, on the other hand, that under the
reasonable assumption that the light quark constituent mass has
sensibly decreased at the cross-over point, the agreement with the
SHM is guaranteed by  $M_s(T=0)\sim 500$ MeV. The limit
$\Lambda\rightarrow\infty$ serves to show that the UV contribution
is not expected to change this picture, even if the analysis in
renormalizable models such as {\em ladder}-QCD ~\cite{bcd} is
necessary.

\subsection{Elementary Collisions}\label{subs:ec}
The location of the cross-over point for EC, $T\sim 160$ MeV
($\mu_B=0$), as well as $M_{i}$, are obtained in the predictive
NJL model with $T_0=170$ MeV in the infinite volume limit.
Consequently, quark multiplicities depend on the volume only
through the explicit dependence of the canonical partition
function on $V,V_c$  ~\cite{noi}. This approximation already
allows us to observe canonical suppression. We find consistency
with our physical picture, as far as processes with zero initial
electric charge are concerned, whereas for pp collisions the Pauli
exclusion principle and canonical suppression balance each other
and $\lambda_S$ remains constant around $\sim 0.3$. In
Fig.~\ref{fig:epem} we plot the predictions of the model compared
to the SHM expectations for \ee collisions.

\section{Conclusions}

We have studied strangeness production within a statistical model
of hadronisation by using an effective model with constituent
quarks. The study supports the idea that the SHM analysis may help
in effective models building for the QCD phase transition, whereas
EM may give a natural explanation of strangeness suppression in
EC, assuming full chemical equilibrium, as due to the absence of
baryon chemical potential and to canonical suppression working in
small volumes. A significative discrepancy is found only in pp
collisions, where further analysis is required.

\newpage
\begin{figure}
\caption{$\lambda_S$ as a function of $\sqrt{s}$ in several kinds
of collisions (from ref.~ \cite{beca1}).\label{fig:ls}}
\end{figure}

\begin{figure}
\caption{Constituent quark masses $M_{q}, M_{s}$ determined by
$\lambda_S \simeq 0.45$ as fitted in Pb-Pb collisions at
$\sqrt{s}_{NN}=17.4$ GeV. The hatched lighter region corresponds
to the free constituent quark gas with $\Lambda=631$ MeV as in
ref.~\cite{njl} ~ whereas the darker hatched region is for
$\Lambda= \infty$. Together, we plot the predictions of the NJL
model with $T_{0}$ ranging from zero (no KMT term) to $\infty$
from left to right (solid line). The black point is for
$T_{0}=170$ MeV. The dashed pointed horizontal and vertical lines
correspond to the $T=\mu_{B}=0$ constituent masses values for the
strange quark and for the light quark $q$ respectively.
\label{bande}}
\end{figure}

\begin{figure}
\caption{Calculated $\lambda_S$ in \ee collisions at $T$=160 MeV
within the NJL model \cite{njl} with $T_0=170$ MeV, as a function
of the total volume $V$ for various single cluster volumes $V_c$.
The dotted line is with $V_c=V$ and the arrow indicates the
grand-canonical limit. The horizontal bands are the ranges of
fitted $\lambda_S$ values in the SHM (see Fig.~\ref{fig:ls}).
\label{fig:epem}}
\end{figure}

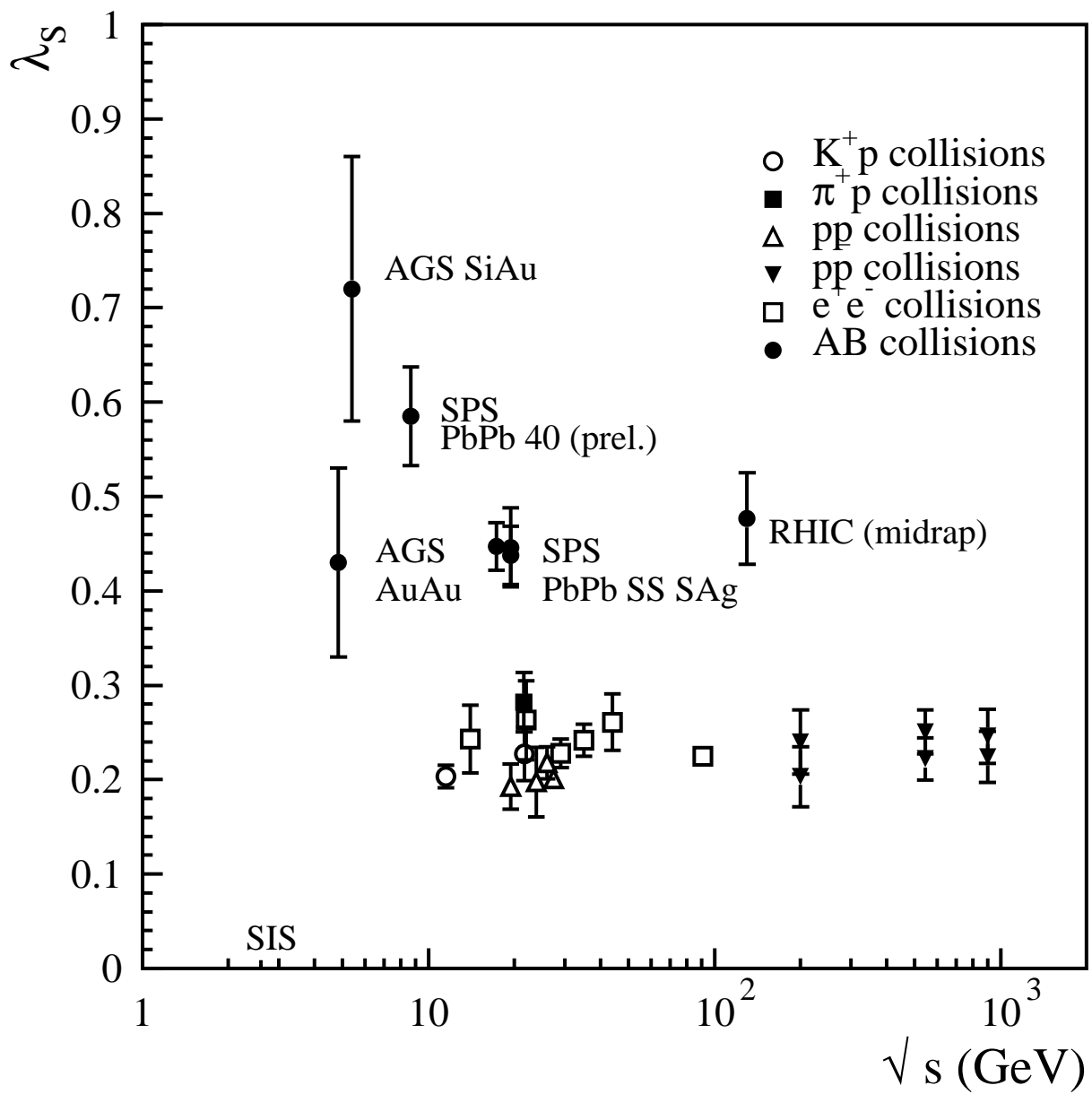
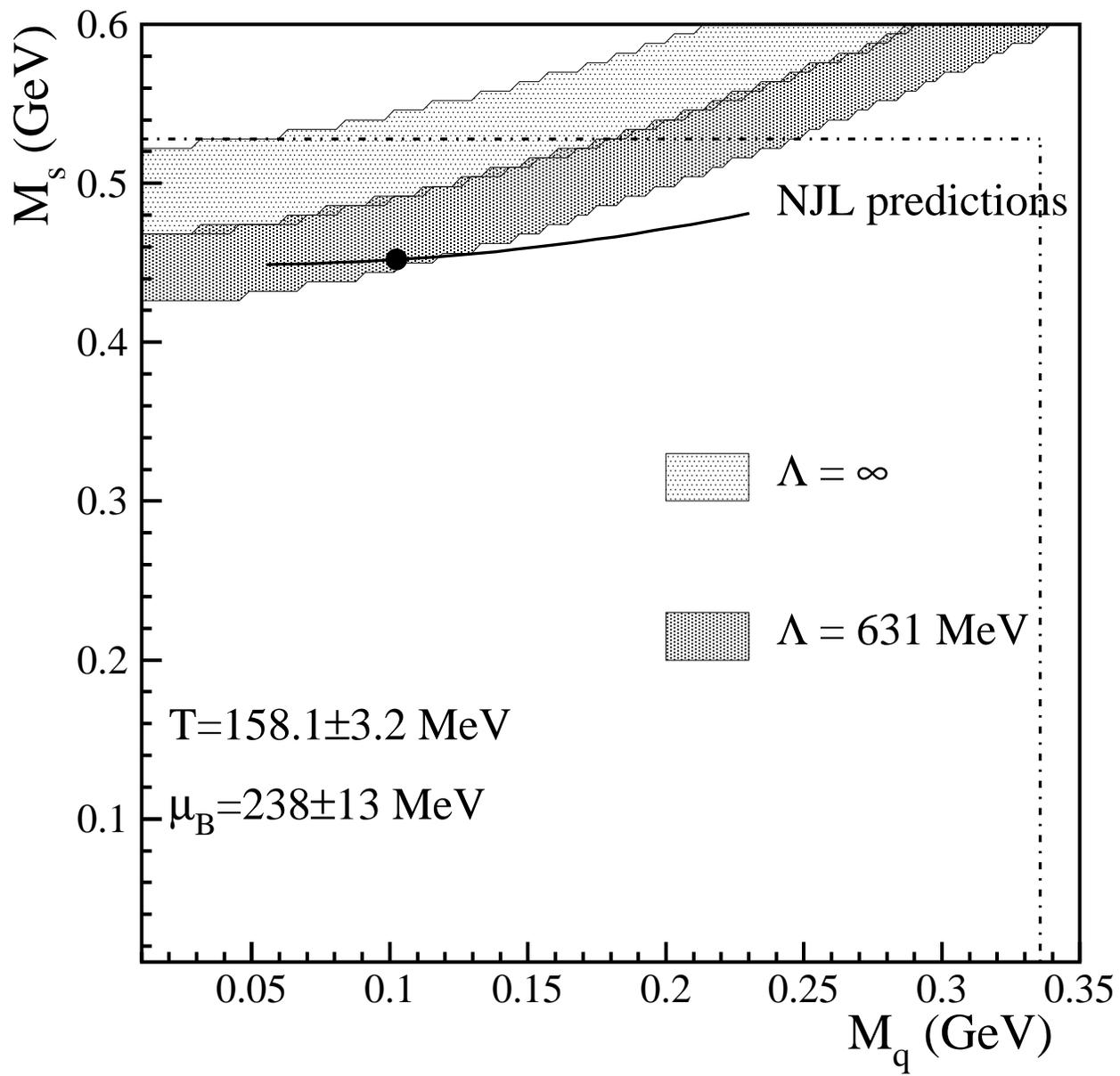
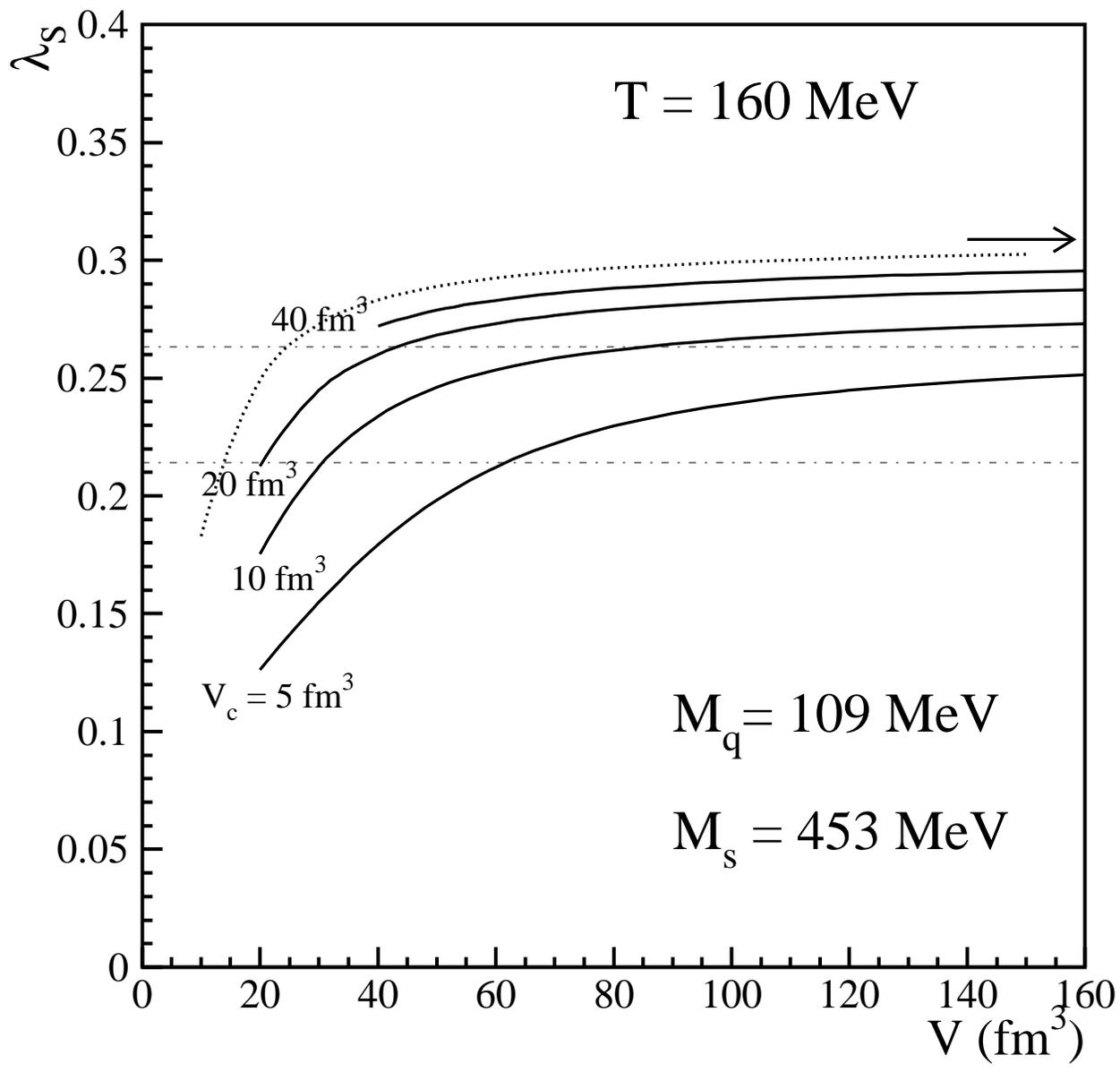
\end{document}